\documentclass[aps,prb,twocolumn,superscriptaddress,amsmath,amsfonts,showpacs]{revtex4}

\usepackage{graphicx}

\begin{document}


\title{
Unambiguous probe of parity-mixing of Cooper pairs in noncentrosymmetric superconductors
}


\author{Satoshi Fujimoto}
\affiliation{Department of Physics, Kyoto University, Kyoto 606-8502, Japan}




\date{\today}

\begin{abstract}
We propose an experimental scheme to detect unambiguously parity-mixing of Cooper pairs in noncentrosymmetric superconductors, which utilizes crossed Andreev reflection processes between two oppositely spin-polarized normal metal leads and a noncentrosymmetric superconductor. It is demonstrated that a non-local conductance exhibits a clear signature of parity breaking of Cooper pairs, and thus, can be a direct probe for
the parity-mixing.
\end{abstract}

\pacs{}


\maketitle



In noncentrosymmetric superconductors (NCSs), which lack inversion symmetry in their crystal structure, antisymmetric spin-orbit (SO) interactions give rise to various exotic effects on superconducting states.\cite{bauer,uir,tog1,Kimura,Sugitani,aki1,ede1,ede2,ede3,gor,Frigeri,fri2,ser,sam2,sam1,yip,yip2,kau,mine,fuji1,haya,fuji2,fuji3,fuji4,yanase,tada,lu,mine2,tanaka,SF}
In particular, pairing states can not be classified according to parity, but, instead, the admixture of spin-singlet pairing and spin-triplet pairing generally occurs.\cite{ede1,gor,Frigeri,fuji2}
This most striking effect, however, unfortunately, has not been detected so far by experimental studies. The difficulty of detecting parity-mixing of Cooper pairs is partly due to the fact that conventional experimental approaches, which are utilized for the determination of parity of Cooper pairs in centrosymmetric superconductors, such as NMR Knight shift measurements, do not provide any useful information concerning parity-mixing. That is, in centrosymmetric superconductors, the change of the Knight shift below the transition temperature $T_c$ tells us whether the pairing state is spin-singlet or spin-triplet. By contrast, in NCS, if the SO interaction is much larger than the energy scale of the superconducting gap, the behavior of the Knight shift below $T_c$ is mainly governed by the strong SO interaction, and does not yield any information on pairing states.\cite{fuji2,fuji3,fri2,yogi,yogi2}
Thus, a novel experimental approach is required for the detection of
parity-mixing of Cooper pairs.
There have been several proposals for this aim, which use, for instance,
tunneling characteristics,\cite{iniotakis,subo,yoko,subo2} 
accidental gap-node structures,\cite{eremin} and 
a fractional vortex scenario,\cite{IFS} etc.

\begin{figure}
\includegraphics[width=7.5cm]{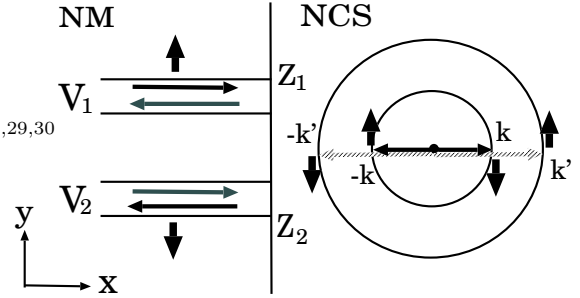}
\caption{Setup for crossed Andreev reflection between a NCS and two 
spatially-separated normal metal (NM) leads with opposite spin polarization
and bias voltages $V_1$ and $V_2$, respectively.
Short arrows represent the directions of electron spins. In the NCS side, 
the circular Fermi surface
split by the Rashba SO interaction is depicted.
Cooper pairs are formed within the same band.
Black (gray) long arrows in the leads 
represent the flow of injected electrons and reflected
holes for the process A (B).
}
\end{figure}

In this paper, we propose another method to probe parity-mixing of Cooper pairs in an unambiguous way. In this scheme, we utilize crossed Andreev reflection (CARE) 
between two ferromagnetic normal metal leads  and a NCS.
CARE is a non-local reflection process; an electron injected from one lead to
a superconductor is converted to a hole in another lead, when
the distance between two leads is smaller than 
the coherence length $\xi$.\cite{CARE1,CARE2,CARE3,CARE4,benja,CARE5}
To illustrate the basic idea, we consider a setup consisting of two 
oppositely spin-polarized leads 
and a NCS with the Rashba SO interaction \cite{ras}
$H_{\rm SO}=\alpha\mbox{\boldmath $\mathcal{L}$}(k)\cdot\mbox{\boldmath $\sigma$}$
with $\mbox{\boldmath $\mathcal{L}$}(k)=(k_y,-k_x,0)$, and $\alpha$ 
the SO coupling constant, 
$\mbox{\boldmath $\sigma$}=(\sigma_x,\sigma_y,\sigma_z)$ 
the Pauli matrices (FIG.1).
In the NCS, the SO interaction splits the electron band into two parts,
$\varepsilon_{k\pm}=\varepsilon_k\pm\alpha |k|$, each of which is
the eigen state of spin chirality.
We assume that the SO split is sufficiently larger than the superconducting gap,
and that there are only intra-band Cooper pairs formed between electrons
in the same band.
For instance, in the $\varepsilon_{k-}$ band, an electron with momentum $k$ and 
spin $\uparrow$ and an electron with momentum $-k$ and spin $\downarrow$ form
a Cooper pairing state $|k \uparrow \rangle |-k \downarrow\rangle$, 
while in the $\varepsilon_{k+}$ band, 
 an electron with momentum $k$ and 
spin $\downarrow$ and an electron with momentum $-k$ and spin $\uparrow$ form
a Cooper pairing state $|k \downarrow \rangle |-k \uparrow\rangle$.
Here, we have chosen the spin quantization axis parallel to 
$\mbox{\boldmath $\mathcal{L}$}(k)=(k_y,-k_x,0)$.
When the density of states in each band is different from each other,
the superposition between these two pairing state is impossible.
Then, the pairing state in each band is
the admixture of a spin-singlet state and a spin-triplet state; 
e.g. $|k \uparrow \rangle |-k \downarrow\rangle=
\frac{1}{2}(|k \uparrow \rangle |-k \downarrow\rangle
-|k \downarrow \rangle |-k \uparrow\rangle)+
\frac{1}{2}(|k \uparrow \rangle |-k \downarrow\rangle +
|k \downarrow \rangle |-k \uparrow\rangle)$.\cite{ede1,gor,Frigeri,fuji2,fuji3}
The superconducting gap of the parity-mixed pairing state is generally given by
$\Delta_{\sigma\sigma'} (k)=\Delta_s(k)i(\sigma_y)_{\sigma\sigma'}
+\mbox{\boldmath $d$}(k)\cdot 
(\mbox{\boldmath $\sigma$}i\sigma_y)_{\sigma\sigma'}$.
We assume that the $\mbox{\boldmath $d$}$-vector of the triplet pairing is 
constrained by the Rashba SO interaction; i.e. 
$\mbox{\boldmath $d$}(k)=|\mbox{\boldmath $d$}(k)|
(\sin \phi_k,-\cos \phi_k,0)$ 
with $\phi_k=\tan^{-1}(k_y/k_x)$.\cite{ede1,gor,Frigeri}
Then, the superconducting gaps on the $\varepsilon_{k+}$-band and 
the $\varepsilon_{k-}$-band are, respectively, 
$\tilde{\Delta}_{+}(k)=[\Delta_s(k)+|\mbox{\boldmath $d$}(k)|]t_k$ and 
$\tilde{\Delta}_{-}(k)=[\Delta_s(k)-|\mbox{\boldmath $d$}(k)|]t_k^{*}$,
where $t_k$ is an odd-parity phase factor given by $t_k=ie^{-i\phi_k}$.
Actually, the structure of the $\mbox{\boldmath $d$}$-vector is determined 
not only by the SO interaction, but also by 
the detail of the pairing interaction.\cite{yanase,tada2}
The generalization of our argument
to the case with more complex structure of $\mbox{\boldmath $d$}$-vector
is straightforward. 
In the following, for simplicity, 
we focus on one-dimensional (1D) scattering problem
where currents flow only along the $x$-axis which is
perpendicular to the interface between the leads and the NCS (see FIG.1). 
A qualitative feature which is important for the detection of parity-mixing
is not largely affected by this simplification.
In fact, the degrees of freedom along the $z$-axis are irrelevant
for our argument.
Effects of titling alignment of the leads on the $xy$-plane will
be discussed later.
In this setup, we assume that
two leads are oppositely spin-polarized with 
the spin-quantization axis parallel to the $y$-axis.

An important observation here is that 
the parity-broken structure of the Cooper pair
$|k \uparrow \rangle |-k \downarrow\rangle$ 
(or $|k \downarrow \rangle |-k \uparrow\rangle$) 
is directly related to the parity-mixing, as explained above.
The parity breaking of Cooper pairs can be detected by CARE
as asymmetric reflection processes;
a process in which an injected electron
with spin $\uparrow$ in the lead 1 is converted to a hole with spin $\downarrow$ 
in the lead 2
is not equivalent to a process in which an injected electron
with spin $\downarrow$ in the lead 2 is converted to a hole with spin $\uparrow$
in the lead 1 because of broken inversion symmetry of the NCS.
In the former process (denoted as the process A), 
the Andreev-reflected hole is associated with
the superconducting gap $\Delta_{-}$, while, in the latter process (denoted as
the process B), the relevant superconducting gap is $\Delta_{+}$.
As mentioned above, in the parity-mixed pairing state,
the amplitudes of these two gaps are different, which can be clearly
observed as a characteristic bias-voltage-dependence of 
the non-local conductance.
Thus, the parity-mixing of Cooper pairs can be detected directly 
without ambiguity.

The non-local conductance, which characterizes the CARE,
is given by
$G_{12}(V_1)=dI_2/dV_1$ and $G_{21}(V_2)=dI_1/d V_2$, where
$I_{1(2)}$ and $V_{1(2)}$ are, respectively, a current and a bias voltage
in the lead 1 (2).
The non-local conductance is expressed in terms of the reflection probabilities
$A_{ij}^{\sigma}$ for the process that an injected electron with spin $\sigma$ 
in the lead $i$ is converted to a hole with spin $-\sigma$ in the lead $j$; i.e.
when the electron spin in each lead is fully polarized,
\begin{eqnarray}
G_{12}(V_1)=G_NA_{12}^{\uparrow}/2, \qquad
G_{21}(V_2)=G_NA_{21}^{\downarrow}/2.
\label{cond}
\end{eqnarray}
Here $G_N$ is the conductance in the normal state.
To obtain the probabilities,
we solve the Bogoliubov-de Gennes (BdG) equation for the CARE.
In the representation where $\sigma_y$ is diagonal, 
the Hamiltonian is decoupled into two parts $\hat{H}_{+}$ and $\hat{H}_{-}$,
each of which corresponds to the pairing state in one of two SO split bands.
The BdG equation for the 1D scattering problem depicted in FIG.1 is
\begin{eqnarray}
\hat{H}_{\nu}\Psi_{\nu}=E\Psi_{\nu} 
\end{eqnarray}
\begin{eqnarray}
\hat{H}_{\nu}=\left(
\begin{array}{cc}
\varepsilon(\hat{k})+\nu\alpha\hat{k}_x+V(x) & -i\nu\Delta_{\nu}(\hat{k}) \\
i\nu\Delta_{\nu}(\hat{k}) & -\varepsilon(\hat{k})-\nu\alpha\hat{k}_x-V(x)
\end{array}
\right),
\end{eqnarray}
with $\nu=\pm$.
Here $\varepsilon(\hat{k})=-\frac{1}{2m}\nabla^2-\mu$ with 
$\mu$ a chemical potential, and $\hat{k}_x=-i\partial_x$.
The gap functions are 
$\Delta_{\pm}(\hat{k})=\Delta_s(\hat{k})\pm |\mbox{\boldmath $d$}(\hat{k})|$.
We assume that a barrier at the interface between 
the lead 1 (2) and the NCS is given by a Dirac-type potential, 
$V(x)=\frac{k_F}{m}Z_{1(2)}\delta(x)$. Here $Z_{1}$ and $Z_2$ are dimensionless 
parameters for
the strength of the barrier potentials, and $k_F$ is defined by $k_F^2/2m=\mu$:
i.e. the Fermi momentum in the case without the SO split.
To simplify the analysis, we consider the case of an $s+p$ wave pairing state,
and neglect $k$-dependence of $\Delta_{\pm}$. The following argument can
be easily extended to the case with more general pairing states such as
a $d+f$ wave state, a $g+h$ wave state etc.
After a straightforward calculation,\cite{btk} 
we obtain the probability $A_{ij}^{\sigma}$,
\begin{eqnarray}
A_{12}^{\uparrow}(E)=\frac{\Delta^2_{\nu}s_{\nu}^2}
{4E^2|\gamma_{\nu}(E)|^2},
\label{ar}
\end{eqnarray}
with $\nu=-$ and $s_{\nu}=1-\nu m\alpha/k_F$. Here, for $E<\Delta_{\nu}$
\begin{eqnarray}
&&4E^2|\gamma_{\nu}(E)|^2=(s_{\nu}E-(Z_1-Z_2)\sqrt{\Delta_{\nu}^2-E^2})^2 
\nonumber \\
&&+(\sqrt{\Delta_{\nu}^2-E^2}(s_{\nu}+2Z_1Z_2)+s_{\nu}(Z_1-Z_2)E)^2,
\end{eqnarray}
and for $E>\Delta_{\nu}$,
\begin{eqnarray}
&&4E^2|\gamma_{\nu}(E)|^2=(s_{\nu}E+(s_{\nu}+2Z_1Z_2)\sqrt{E^2-\Delta^2_{\nu}})
)^2 \nonumber \\
&&+(Z_1-Z_2)^2(s_{\nu}E+\sqrt{E^2-\Delta^2_{\nu}})^2.
\end{eqnarray}
The probability $A_{12}^{\downarrow}(E)$, 
is given by Eq.(\ref{ar})
with $\nu=+$.
One can obtain the probability $A_{21}^{\sigma}$  
by interchanging $Z_1$ and $Z_2$ in the expression of $A_{12}^{\sigma}$.
In the derivation of $A_{ij}^{\sigma}$, we have used the approximation
that the Fermi momentum for $\varepsilon_{k\pm}$ is 
$k_{F\pm}\approx k_F\mp m\alpha$, which is valid when the SO split is 
much smaller than the Fermi energy.
In fact, within this approximation, 
the shift of the Fermi momentum due to the SO split does not change qualitatively
the feature of the non-local conductance 
that is important for the detection of parity-mixing of Cooper pairs,
as will be shown below.

We, first, consider the case that the leads 1, 2 are fully spin-polarized
in the opposite directions,
and the non-local conductance is given by (\ref{cond}).
Although the spin polarization in the leads
induces exchange fields in the NCS region 
which may affect the amplitude of the superconducting gap
in a nontrivial way, we neglect this effect because it may not change
our argument qualitatively.
Because of the parity-broken structure of Cooper pairs mentioned above,
$G_{12}(eV)$ and $G_{21}(eV)$ exhibit asymmetric behaviors as functions of
$V$ even when
$Z_1=Z_2$; i.e.
for $G_{12}(eV)$ a peak structure appears at $eV=\Delta_{-}$, while 
for $G_{21}(eV)$ it appears at $eV_{2}=\Delta_{+}$.
When there is the admixture of spin-singlet pairing
and spin-triplet pairing, $\Delta_{+}\neq\Delta_{-}$ holds.
Thus, the parity-mixing can be detected unambiguously from
the measurement of the non-local conductance.
Also, we can derive 
the BCS gap magnitudes for the spin-singlet pairs and spin-triplet pairs 
from $\Delta_s=(\Delta_{+}+\Delta_{-})/2$, and
$|\mbox{\boldmath $d$}(k)|=(\Delta_{+}-\Delta_{-})/2$.
Typical behaviors of the non-local conductance as functions of bias voltages
are shown in FIG.2(a).
It is noted that even in the case of $Z_1\neq Z_2$, 
the origin of this asymmetric behavior of $G_{12}$ and 
$G_{21}$ can be clearly attributed to the result of parity-mixing,
since the most important factor which yields the asymmetric behavior of 
the non-local conductance is the existence of two gaps 
$\Delta_{+}$ and $\Delta_{-}$ associated with, respectively, opposite spin
chirality of the two SO split Fermi surfaces.
We emphasize that the two different gap structure which appears 
in $G_{12}$ and $G_{21}$ shown in FIG.2(a)
is obviously different from 
conventional multi-gap behaviors of centrosymmetric 
superconductors with multi-bands.
From this point of view, the CARE experiment is more advantageous
than the conventional Andreev reflection experiment \cite{iniotakis,subo,yoko,subo2}
as a probe for parity-mixing, though its realization is
still challenging with current nanotechnology.
In the above argument, a crucial assumption for the gap function is that
there are no inter-band Cooper pairs, or, if they exist, the gap amplitude
for the inter-band pairs is negligibly small.
This assumption is valid as long as the SO split of the Fermi surface is
sufficiently smaller than the Fermi energy.

\begin{figure}
\includegraphics[width=6.5cm]{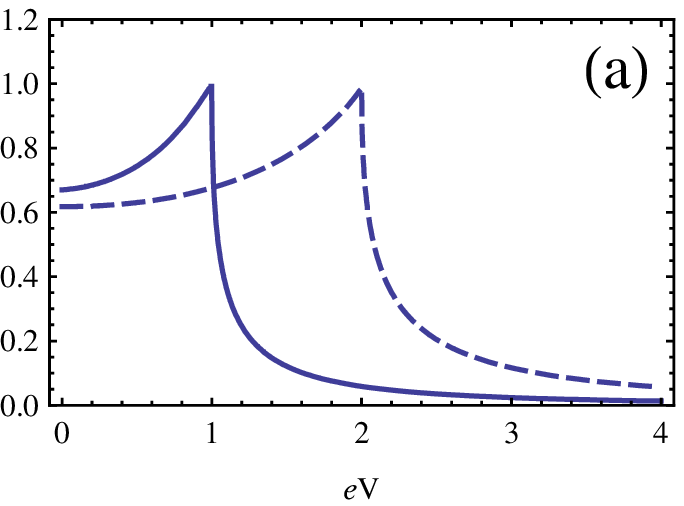}
\includegraphics[width=6.5cm]{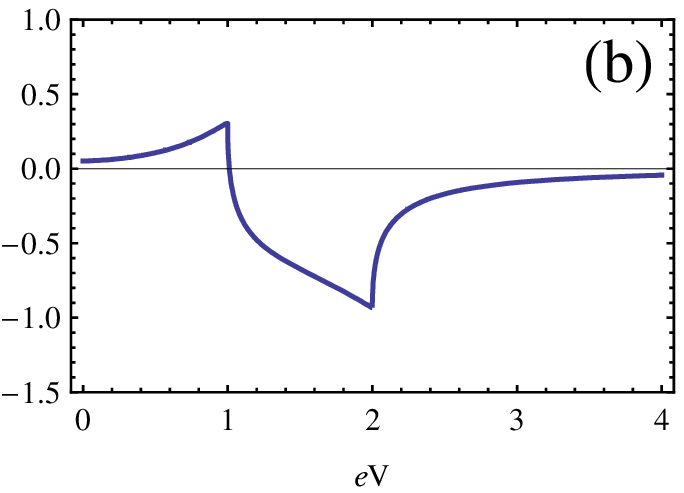}
\caption{(a) Typical behaviors of 
the non-local conductance as functions of a bias voltage $eV$.
$G_{12}(eV)$ (solid line) and $G_{21}(eV)$ (broken line) for
$\Delta_{-}=1.0$ (energy unit), $\Delta_{+}=2.0$, 
$Z_1=0.3$, $Z_2=0.4$, $m\alpha/k_F=0.1$.
Unit of the conductance is $G_N/2$.
(b) $(G_{12}^{H>0}-G_{12}^{H<0})/(1-C_0)$ as a function of a bias voltage
for the same parameters as (a).
}
\end{figure}

An apparent drawback of the above scheme is that in the setup shown in FIG.1,
one needs to know beforehand the spin structure on the Fermi surfaces of the NCS
determined by the SO interaction
to align the direction of the spin polarization in two leads properly. 
Generally, this task is not so easy, because the structure of the SO interaction
for real NCSs is a quite complicated function of 
momentum $k$.\cite{sam3,fermi1,fermi2}  
For the purpose of extending our scenario to such realistic cases with 
the non-Rashba SO interactions, 
we consider a setup a bit different from that depicted in FIG.1; 
in two normal metal leads,
there is no spontaneous magnetization, but, instead, 
spin polarization is induced by an external weak magnetic field $H$ which is
smaller than $H_{c1}$.
We neglect effects of the magnetic field 
on the NCS, since it does not change our argument qualitatively.
A main effect of a sufficiently small magnetic field on the Andreev reflection
 processes is to raise an imbalance of spin population in two normal metal leads.
Suppose the Rashba SO interaction for a while.
We will discuss more general cases later.
Then, when the magnetic field is parallel to 
the positive direction of the $y$-axis, the non-local conductance $G_{12}^{H>0}$ is
\begin{eqnarray}
\frac{2G_{12}^{H>0}}{G_N}=A_{12}^{\uparrow}+C_0A_{12}^{\downarrow},
\label{gh}
\end{eqnarray}
where $C_0=N_{\downarrow}/N_{\uparrow}$ 
with $N_{\uparrow(\downarrow)}$ the total number of electrons with up (down) spin
in the lead 1, and $A_{12}^{\uparrow(\downarrow)}$ in Eq.(\ref{gh}) 
is the reflection probability
in the case without a magnetic field.
In the derivation of (\ref{gh}), 
we have taken into account the Zeeman effect up to
the lowest order in $\mu_{\rm B}H/\mu$, but
neglected the change of the Fermi momentum due to 
the Zeeman shift in the leads.
Thus, the suppression of the Andreev reflection due to
the spin polarization is not included in Eq.(\ref{gh}) 
up to $O(\mu_{\rm B}H/\mu)$.
This approximation does not affect an important feature of the non-local 
conductance relevant to the detection of parity-mixing for a sufficiently small
$H$, as will be clarified later.
In a similar way,
the non-local conductance in the case with 
a magnetic field parallel to the negative direction of the $y$-axis
is obtained as
$G_{12}^{H<0}/G_N=C_0A_{12}^{\uparrow}+A_{12}^{\downarrow}$. From 
the difference between the conductance for $H>0$ and that for $H<0$, 
$G^{H>0}_{12}-G^{H<0}_{12}=
(1-C_0)(A_{12}^{\uparrow}-A_{12}^{\downarrow})$, we can clearly see 
whether the parity of Cooper pairs is broken 
($A_{12}^{\uparrow}\neq A_{12}^{\downarrow}$) or not 
($A_{12}^{\uparrow}= A_{12}^{\downarrow}$).
We show a typical behavior of $(G^{H>0}_{12}-G^{H<0}_{12})/(1-C_0)$
as a function of a bias voltage in FIG.2(b).
This quantity exhibits distinct peak structures 
at $eV=\Delta_{+}$ and $\Delta_{-}$
as a signature of the parity-mixing, and thus, can be a useful probe
for the admixture of spin-singlet pairs and spin-triplet pairs.
It is noted that the peak height at $eV=\Delta_{\pm}$ is not affected by
the Zeeman shift of the Fermi momentum in the NM leads up to
the first order in $\mu_{\rm B}H/\mu$, since, up to this order,
the magnetic field $H$ enters into the expression of $A_{ij}^{\sigma}$ 
in the form of $(\mu_{\rm B}H/\mu)\sqrt{|\Delta_{\pm}^2-E^2|}$.
Thus, the approximation used in the derivation of Eq.(\ref{gh}) and
the equation for $G_{12}^{H<0}$ is valid for our purpose.
This scheme which uses field-induced spin polarization in the leads 
can be utilized for
the detection of parity-mixing in the general case
that the structure of the antisymmetric SO interaction, 
$H_{\rm SO}=
\alpha\mbox{\boldmath $\mathcal{L}$}(k)\cdot\mbox{\boldmath $\sigma$}$, 
is unknown.
Even in this case, when a magnetic field is applied,
one can observe
the asymmetry between the non-local conductance for a magnetic field
with a certain direction, $G_{12}^{H>0}$, and the non-local conductance for
a field anti-parallel to it, $G_{12}^{H<0}$,
quite generally except in the case with 
$\vec{H}\perp\mbox{\boldmath $\mathcal{L}$}(k)$.
Thus, it is not difficult to find a direction of the magnetic field
for which the asymmetric behavior of the non-local conductance is observed,
and the conductance difference $G^{H>0}_{12}-G^{H<0}_{12}$ is nonzero.
Then, one can detect parity-mixing of Cooper pairs.

Finally, we comment on effects of inversion symmetry breaking caused by the
interface between the NCS and the NM leads.
Generally, inversion symmetry is broken at a surface.
However, this extrinsic inversion symmetry breaking does not affect our proposal
because of the following reason.
For the $(100)$-interface depicted in FIG.1, the SO interaction due to
the interface is typically the Rashba-type with the Hamiltonian, 
$H'=\alpha'(k_z\sigma_y-k_y\sigma_z)$.
On the other hand, CARE processes for this geometry are dominated by
electrons and holes with momentum parallel to the $x$-axis, i.e. $k_y=k_z=0$.
Thus, effects of 
the extrinsic Rashba interaction on
the CARE are negligible.

During the preparation of this manuscript, we have become aware of
the paper by Wu and Samokhin (arXiv:0904.2397), in which 
the conductance for conventional Andreev reflection between a ferromagnetic metal
and a NCS is calculated in a thorough way.
It is important to generalize the current study
to more realistic situations as considered by Wu and Samokhin
for quantitative comparison between theory and
experiments.

In summary, it is proposed that a crossed Andreev reflection experiment
can be utilized as an unambiguous probe for parity-mixing of Cooper pairs
in NCSs. 
 
This work is supported by the Grant-in-Aids for
Scientific Research from MEXT of Japan
(Grant No.18540347, Grant No.19052003).

\end{document}